\documentclass[amsmath,amssymb,showpacs]{revtex4}
\usepackage{epsfig}
\usepackage{graphicx}

\begin{document}

\title{Gravitational deflection of light in Rindler-type potential as a possible resolution to the observations of Bullet Cluster 1E0657-558}

\author{Xin Li}
\email{lixin@ihep.ac.cn}
\author{Zhe Chang}
\email{changz@ihep.ac.cn}
\affiliation{Institute of High Energy Physics\\
Theoretical Physics Center for Science Facilities\\
Chinese Academy of Sciences, 100049 Beijing, China}

\begin{abstract}
The surface density $\Sigma$-map and the convergence $\kappa$-map of Bullet Cluster 1E0657-558 show that the center of baryonic matters separates from the center of gravitational force, and the distribution of gravitational force do not possess spherical symmetry. This hints that a modified gravity with difference to Newtonian inverse-square law at large scale, and less symmetry is worth investigating. In this paper, we study the dynamics in Randers-Finsler spacetime. The Newtonian limit and gravitational deflection of light in a Rindler-type potential is focused in particular. It is shown that the convergence in Finsler spacetime could account for the observations of Bullet Cluster.
\end{abstract}

\maketitle
\section{Introduction}
In 1933, Zwicky\cite{Zwicky} analysed the velocity dispersion for the Coma cluster. His analysis implied that the Coma cluster is full of invisible matters. There are a great variety of observations which show that  the
rotational velocity curves of all spiral galaxies tend to
constant values\cite{Trimble}. These include the Oort\cite{Oort} discrepancy in
the disk of the Milky Way\cite{Bahcall}, the velocity dispersions of
dwarf Spheroidal galaxies\cite{Vogt}, and the flat rotation curves
of spiral galaxies\cite{Rubin}. These facts
 violate sharply the prediction of Newton's inverse square law of gravitation.

The most widely adopted way to resolve these difficulties is the
dark matter hypothesis. It is assumed that all visible stars are
surrounded by massive nonluminous matters. Though it explains the
flat rotation curves of spiral galaxies, the hypothesis has its own
weakness. No theory predicts these matters, and they behave in such
ad hoc way. There are a lot of possible candidates of dark matter
(such as axion, neutrino {\it et al}), but none of them
satisfactory. Up to now, all of them either undetected or excluded
by observation.

Because of these troubles induced by dark matter, some models have
been built for alternative of the dark matter hypothesis. Their main
ideas are to assume that the Newtonian gravity or Newton's dynamics
is invalid in galactic scale. The famous one is Milgrom's Modified
Newtonian Dynamics (MOND)\cite{Milgrom}. MOND assumes that the
Newtonian dynamics does not hold in galactic scale. As
a phenomenological model, MOND explains well the flat rotation
curves with a simple formula and a new parameter. In particular, it
deduces naturally a well-known global scaling relation for spiral
galaxies, the Tully-Fisher relation\cite{TF}. By introducing several
scalar, vector and tensor fields, Bekenstein\cite{Bekenstein}
rewrote the MOND into a covariant formulation. He showed that the
MOND satisfies all four classical tests on Einstein's general
relativity in solar system.

However, MOND still face challenges. The strong and weak gravitational lensing observations of Bullet Cluster 1E0657-558\cite{Bullet} could not be explained well by MOND and Bekenstein's relativistic version of MOND\cite{Angus}. Takahashi {\it et al}. \cite{Takahashi} have investigated weak gravitational lensing of three galaxy clusters ((A1689, CL0024+1654, CL1358+6245) in terms of the MOND. They found that MOND could not explain the data of three galaxy clusters unless a dark matter halo is added. Problems still remain in fitting X-ray temperature of galaxy clusters with MOND. In the framework of MOND, a plausible resolution of these issues is the ``marriage" of MOND with neutrino of mass $m_\nu\geq2$eV\cite{Angus,Takahashi}. However, the WMAP7 data give an upper limit of 1.2eV for the sum of neutrino mass \cite{Hannestad}.

The main feature of the Bullet Cluster is that the magnitude of the gravitational force approximately equals five times of what produced by baryonic matters. Beside the main feature, there are two particular features too. The surface density $\Sigma$-map reconstructed from X-ray imaging observations gives the center of baryonic matters. The convergence $\kappa$-map reconstructed
from strong and weak gravitational lensing observations gives the center of gravitational force. The first particular feature is that the center of baryonic matters separates from the center of gravitational force. The convergence $\kappa$-map manifests that the distribution of gravitational force do not possess spherical symmetry. This is the other particular feature. These two particular features are the reasons why MOND could not explain well the observations of Bullet Cluster.

Due to the reasons given above, the observations of Bullet Cluster have been regarded as a direct evidence of dark matter. The dominance of dark matter has lasted almost six decades. However, up to now, no direct evidence claim that dark matter is detected. It is very interesting to search an alternative to dark matter. One famous and successful model is modified gravity (MOG)\cite{Moffat}. MOG assumes that some non geometrical fields couple to the gravitational field of general relativity. MOG explains well the flat rotation curves of galaxies\cite{Brownstein}. The MOG's prediction for the $\kappa$-map results in two baryonic components distributed across the Bullet Cluster 1E0657-558 with averaged mass-fraction of 83\% intracluster medium (ICM) gas and 17\% galaxies\cite{Brownstein1}. One should notice that the locally measured value of Newton's constant $G$ varies spatially in MOG. In dark matter hypothesis, the galaxy's mass involve baryonic mass and the mass of dark matter. As for MOG, the galaxy's mass only involve baryonic mass, and the effective acceleration depends on the running gravitational coupling $G(r)$.

In this paper, we try to introduce a possible alternative to dark matter, which could account for the observations of Bullet Cluster.
In 1912, A. Einstein proposed his famous general relativity which gives the connection between Riemann geometry and gravitation. In general relativity, the effects of gravitation are ascribed to spacetime curvature instead of a force. By mimicking Einstein, we may investigate the gravity in Finsler spacetime\cite{Book by Bao}. Finsler geometry as a natural generalization of Riemann geometry could provide new sight on modern physics.

The Finsler gravity modifies the Newtonian inverse-square law at large scale. It reduces to the Newtonian inverse-square law at small distance. The center of gravitational force may separate from the center of baryonic matters at large scale. In general, Finsler spacetime admits less Killing vectors than Riemann spacetime. In Finsler spacetime, the spherical symmetry may be broken at large scale. Finsler spacetime possesses the two particular features of Bullet Cluster. It could account for the observations of Bullet Cluster.

Recently, Grumiller \cite{Grumiller} constructed an effective model for gravity of a central object at large scales. In Grumiller's model, to leading order in the large radius expansion, all terms are expected from general relativity, except for
the Rindler one. The Rindler term leads to an anomalous acceleration \cite{Wald}, which could account for the Pioneer anomaly \cite{Anderson} in solar system and the rotational curve of galaxies. In this paper, we will show that the Rindler-type potential could account for the observations of Bullet Cluster within the framework of Finsler gravity.

This paper is organized as follows. In Sec.2, we present the vacuum field equation in Finsler spacetime. In Sec.3, by making use of the post-Newtonian approximation and the viewpoints of Zermelo navigation problem, we investigate the dynamics in Randers-Finsler spacetime. The emphasis is focused on the Newtonian limit and gravitational deflection of light in a Rindler-type potential. The concluding remarks are given is Sec.4.

\section{Vacuum field equation in Finsler spacetime}
Instead of defining an inner product structure over the tangent bundle in Riemann geometry, Finsler geometry is based on
the so called Finsler structure $F$ with the property
$F(x,\lambda y)=\lambda F(x,y)$ for all $\lambda>0$, where $x$ represents position
and $y\equiv\frac{dx}{d\tau}$ represents velocity. The Finsler metric is given as\cite{Book
by Bao}
 \begin{equation}
 g_{\mu\nu}\equiv\frac{\partial}{\partial
y^\mu}\frac{\partial}{\partial y^\nu}\left(\frac{1}{2}F^2\right).
\end{equation}
Finsler geometry has its genesis in integrals of the form
\begin{equation}
\label{integral length}
\int^r_sF(x^1,\cdots,x^n;\frac{dx^1}{d\tau},\cdots,\frac{dx^n}{d\tau})d\tau~.
\end{equation}
The Finsler structure represents the length element of Finsler space.

The parallel transport
has been studied in the framework of Cartan
connection\cite{Matsumoto,Antonelli,Szabo}. The notation of parallel
transport in Finsler manifold means that the length
$F\left(\frac{dx}{d\tau}\right)$ is constant.
The geodesic equation for Finsler manifold is given as\cite{Book by Bao}
\begin{equation}
\label{geodesic}
\frac{d^2x^\mu}{d\tau^2}+G^\mu=0,
\end{equation}
where
\begin{equation}
\label{geodesic spray}
G^\mu=\frac{1}{2}g^{\mu\nu}\left(\frac{\partial^2 F^2}{\partial x^\lambda \partial y^\nu}y^\lambda-\frac{\partial F^2}{\partial x^\nu}\right)
\end{equation} is called geodesic spray coefficient.
Obviously, if $F$ is Riemannian metric, then
\begin{equation}
G^\mu=\tilde{\gamma}^\mu_{\nu\lambda}y^\nu y^\lambda,
\end{equation}
where $\tilde{\gamma}^\mu_{\nu\lambda}$ is the Riemannian Christoffel symbol.
Since the geodesic equation (\ref{geodesic}) is directly
derived from the integral length of $\sigma$
\begin{equation} L=\int
F\left(\frac{dx}{d\tau}\right)d\tau,
\end{equation} the inner product
$\left(\sqrt{g_{\mu\nu}\frac{dx^\mu}{d\tau}\frac{dx^\nu}{d\tau}}=F\left(\frac{dx}{d\tau}\right)\right)$
of two parallel transported vectors is preserved.

In Finsler manifold, there exists a linear connection~-~the
Chern connection\cite{Chern}. It is torsion freeness and almost
metric-compatibility,
 \begin{equation}\label{Chern connection}
 \Gamma^{\alpha}_{\mu\nu}=\gamma^{\alpha}_{\mu\nu}-g^{\alpha\lambda}\left(A_{\lambda\mu\beta}\frac{N^\beta_\nu}{F}-A_{\mu\nu\beta}\frac{N^\beta_\lambda}{F}+A_{\nu\lambda\beta}\frac{N^\beta_\mu}{F}\right),
 \end{equation}
 where $\gamma^{\alpha}_{\mu\nu}$ is the formal Christoffel symbols of the
second kind with the same form of Riemannian connection, $N^\mu_\nu$
is defined as
$N^\mu_\nu\equiv\gamma^\mu_{\nu\alpha}y^\alpha-A^\mu_{\nu\lambda}\gamma^\lambda_{\alpha\beta}y^\alpha
y^\beta$
 and $A_{\lambda\mu\nu}\equiv\frac{F}{4}\frac{\partial}{\partial y^\lambda}\frac{\partial}{\partial y^\mu}\frac{\partial}{\partial y^\nu}(F^2)$ is the
 Cartan tensor (regarded as a measurement of deviation from the Riemannian
 Manifold). In terms of Chern connection, the curvature of Finsler space is given as
\begin{equation}\label{Finsler curvature}
R^{~\lambda}_{\kappa~\mu\nu}=\frac{\delta
\Gamma^\lambda_{\kappa\nu}}{\delta x^\mu}-\frac{\delta
\Gamma^\lambda_{\kappa\mu}}{\delta
x^\nu}+\Gamma^\lambda_{\alpha\mu}\Gamma^\alpha_{\kappa\nu}-\Gamma^\lambda_{\alpha\nu}\Gamma^\alpha_{\kappa\mu},
\end{equation}
where $\frac{\delta}{\delta x^\mu}=\frac{\partial}{\partial x^\mu}-N^\nu_\mu\frac{\partial}{\partial y^\nu}$.

The gravity in Finsler spacetime has been investigated for a long
time\cite{Takano,Ikeda,Tavakol1, Bogoslovsky1}. In this paper, we introduce vacuum field equation from the extension of an analogy, which was discussed first by Pirani\cite{Pirani, Rutz}. In Newton's theory of gravity, the equation of motion of a test particle is given as
\begin{equation}
\label{dynamic Newton}
\frac{d^2x^i}{dt^2}=-\eta^{ij}\frac{\partial \phi}{\partial x^i},
\end{equation}
where $\phi=\phi(x)$ is the gravitational potential and $\eta^{ij}$ is Euclidean metric. For an infinitesimal transformation $x^i\rightarrow x^i+\epsilon\xi^i$($|\epsilon|\ll1$), the equation (\ref{dynamic Newton}) becomes, up to first order in $\epsilon$,
\begin{equation}
\label{dynamic Newton1}
\frac{d^2x^i}{dt^2}+\epsilon\frac{d^2\xi^i}{dt^2}=-\eta^{ij}\frac{\partial \phi}{\partial x^i}-\epsilon\eta^{ij}\xi^k\frac{\partial^2\phi}{\partial x^j\partial x^k}.
\end{equation}
Combining the above equations(\ref{dynamic Newton}) and (\ref{dynamic Newton1}), we obtain
\begin{equation}
\frac{d^2\xi^i}{dt^2}=\eta^{ij}\xi^k\frac{\partial^2\phi}{\partial x^j\partial x^k}\equiv\xi^kH^i_k.
\end{equation}
In Newton's theory of gravity, the vacuum field equation is given as $H^i_i=\bigtriangledown^2\phi=0$. It means that the tensor $H^i_k$ is traceless in Newton's vacuum.

In general relativity, the geodesic deviation gives similar equation
\begin{equation}
\frac{D^2\xi^\mu}{D\tau^2}=\xi^\nu H^\mu_\nu,
\end{equation}
where $H^\mu_\nu=\tilde{R}^{~\mu}_{\lambda~\nu\rho}\frac{dx^\lambda}{d\tau}\frac{dx^\rho}{d\tau}$. Here, $\tilde{R}^{~\mu}_{\lambda~\nu\rho}$ is Riemannian curvature tensor, $D$ denotes the covariant derivative alone the curve $x^\mu(\tau)$. The vacuum field equation in general relativity gives $\tilde{R}_{\mu\nu}=\tilde{R}^{~\mu}_{\lambda~\lambda\nu}=0$\cite{Weinberg}. It implies that the tensor $H^\mu_\nu$ is also traceless, $H=H^\mu_\mu=0$.

In Finsler spacetime, the geodesic deviation gives\cite{Book by Bao}
\begin{equation}
\frac{D^2\xi^\mu}{D\tau^2}=\xi^\nu H^\mu_\nu,
\end{equation}
where $H^\mu_\nu=R^{~\mu}_{\lambda~\nu\rho}\frac{dx^\lambda}{d\tau}\frac{dx^\rho}{d\tau}$. Here, $R^{~\mu}_{\lambda~\nu\rho}$ is Finsler curvature tensor defined in (\ref{Finsler curvature}), $D$ denotes covariant derivative $\frac{D\xi^\mu}{D\tau}=\frac{d\xi^\mu}{d\tau}+\xi^\nu\frac{dx^\lambda}{d\tau}\Gamma^\mu_{\nu\lambda}(x,\frac{dx}{d\tau})$. Since the vacuum field equations of Newton's gravity and general relativity have similar form, we may assume that vacuum field equation in Finsler spacetime holds similar requirement as the case of Netwon's gravity and general relativity. It implies that the tensor $H^\mu_\nu$ in Finsler geodesic deviation equation should be traceless, $H=0$.

The notion of Ricci tensor in Finsler geometry was first introduced by Akbar-Zadeh\cite{Akbar}
\begin{equation}
Ric_{\mu\nu}=\frac{\partial^2\left(\frac{1}{2}F^2 R\right)}{\partial y^\mu\partial y^\nu},
\end{equation}
where $R=\frac{y^\mu}{F}R^{~\kappa}_{\mu~\kappa\nu}\frac{y^\nu}{F}$. And the scalar curvature in Finsler geometry is given as $S=g^{\mu\nu}Ric_{\mu\nu}$. Constructing a physical Finslerian theory of gravity in arbitrary Finsler spacetime is a difficult task.
However, it has been pointed out that constructing a Finslerian theory of gravity in Finlser spacetime of Berwald type is viable \cite{Tavakol}. A Finsler spacetime is said to be of Berwald type if the Chern connection (\ref{Chern connection}) has no $y$ dependence\cite{Book by Bao}. In light of the research of Tavakol {\it et al}. \cite{Tavakol}, the gravitational field equation in Berwald-Finsler space has been studied in Ref. \cite{Finsler DM,Lixin}. In Berwald-Finsler space, the Ricci tensor reduces to
\begin{equation}
Ric_{\mu\nu}=\frac{1}{2}(R^{~\alpha}_{\mu~\alpha\nu}+R^{~\alpha}_{\nu~\alpha\mu}).
\end{equation}
It is manifestly symmetric and covariant. Apparently the Ricci
tensor will reduce to the Riemann-Ricci tensor if the metric tensor $g_{\mu\nu}$ does not depend on $y$. We starts from the
second Bianchi identities on Berwald-Finsler space\cite{Book by Bao}
\begin{equation}\label{Bianchi
on Riemann} R^{~\alpha}_{\mu~\lambda\nu|\beta}+R^{~\alpha}_{\mu~\nu
\beta|\lambda}+R^{~\alpha}_{\mu~\beta\lambda|\nu}=0,
\end{equation} where the
$|$ means the covariant derivative. The metric-compatibility
\begin{equation}
g_{\mu\nu|\alpha}=0~~~~\mathrm{and}~~~~g^{\mu\nu}_{~~|\alpha}=0,
\end{equation}
and contraction of (\ref{Bianchi on Riemann}) with $g^{\mu\beta}$
gives that
\begin{equation} R^{\mu
\alpha}_{~~\lambda\nu|\mu}+R^{\mu\alpha}_{~\nu\mu|\lambda}+R^{\mu\alpha}_{~\mu\lambda|\nu}=0.
\end{equation}
Lowering the index $\alpha$ and contracting with
$g^{\alpha\lambda}$,  we obtain
\begin{equation}
\left[Ric_{\mu\nu}-\frac{1}{2}g_{\mu\nu}S\right]_{|\mu}+\left\{\frac{1}{2}B^{~\alpha}_{\alpha~\mu\nu}+B^{~\alpha}_{\mu~\nu\alpha}\right\}_{|\mu}=0
,\end{equation} where
\begin{equation}
B_{\mu\nu\alpha\beta}=-A_{\mu\nu\lambda}R^{~\lambda}_{\theta~\alpha\beta}y^\theta/F.
\end{equation}
 Thus, the counterpart of the Einstein's field equation on
 Berwald~-~Finsler space  takes the form
 \begin{equation}\label{berwald field eq}
\left[Ric_{\mu\nu}-\frac{1}{2}g_{\mu\nu}S\right]+\left\{\frac{1}{2}B^{~\alpha}_{\alpha~\mu\nu}+B^{~\alpha}_{\mu~\nu\alpha}\right\}=8\pi
G T_{\mu\nu}.
\end{equation}
In Eq. (\ref{berwald field eq}), the term in ``[~]" is symmetrical tensor, and the term in ``\{\}" is asymmetrical tensor.
By making use of the equation (\ref{berwald field eq}), the vacuum field equation in Finsler spacetime of Berwald type implies
\begin{equation}
Ric_{\mu\nu}=\frac{1}{2}(R^{~\alpha}_{\mu~\alpha\nu}+R^{~\alpha}_{\nu~\alpha\mu})=0.
\end{equation}
It means that the tensor $H^\mu_\nu=R^{~\mu}_{\lambda~\nu\rho}\frac{dx^\lambda}{d\tau}\frac{dx^\rho}{d\tau}$ is traceless in Finsler spacetime of Berwald type. Therefore, the analogy from the geodesic deviation equation is valid at least in Finsler spacetime of Berwald type. For this reason, we may suppose that this analogy is valid in general Finsler spacetime.
For the rest of the paper, we adopt $H=0$ as a requirement of vacuum field equation in Finsler spacetime.

\section{The dynamics in Randers-Finsler spacetime}
\subsection{The post-Newtonian approximation in Randers-Finsler spacetime}
The Randers space is a special kind of Finsler geometry with Finsler
structure $F$  on the slit tangent bundle $TM\backslash0$ of a
manifold $M$,
\begin{equation}
F(x,y)\equiv \alpha(x,y)+\beta(x,y),
\end{equation} where
\begin{eqnarray}
\alpha (x,y)&\equiv&\sqrt{\tilde{a}_{\mu\nu}(x)y^\mu y^\nu},\\
\beta(x,y)&\equiv& \tilde{b}_\mu(x)y^\mu,
\end{eqnarray} and $\tilde{a}_{ij}$ is Riemannian metric.

The geodesic spray coefficient $G^\mu$ (\ref{geodesic spray}) in Randers-Finsler spacetime reads\cite{Book by Bao}
\begin{eqnarray}
G^\mu=(\tilde{\gamma}^\mu_{\nu\lambda}+l^\mu\tilde{b}_{\nu|\lambda})y^\nu y^\lambda+(\tilde{a}^{\mu\nu}-l^\mu\tilde{b}^\nu)(\tilde{b}_{\nu|\lambda}-\tilde{b}_{\lambda|\nu})\alpha\left(\frac{dx}{d\tau}\right) y^\lambda,
\end{eqnarray}
where $l^\mu\equiv y^\mu/F$, $\tilde{\gamma}^\mu_{\nu\lambda}$ is the Christoffel symbol of Riemannian metric $\tilde{a}$ and $\tilde{b}_{\nu|\lambda}$ denotes the covariant derivative with respect to the Riemannian metric $\tilde{a}$
\begin{equation}
\tilde{b}_{\nu|\lambda}=\frac{\partial\tilde{b}_\nu}{\partial x^\lambda}-\tilde{\gamma}^\mu_{\nu\lambda}\tilde{b}_\mu.
\end{equation}
In the rest of the paper, we just consider the case that $\beta$ is a closed 1-form. Thus, the geodesic equation of such Randers spacetime is given as
\begin{equation}
\frac{d^2x^\mu}{d\tau^2}+(\tilde{\gamma}^\mu_{\nu\lambda}+l^\mu\tilde{b}_{\nu|\lambda})y^\nu y^\lambda=0.
\end{equation}
In post-Newtonian approximation, to first order ($GM/r$), the non vanish Christoffel symbols of Riemannian metric $\tilde{a}$
are $\tilde{\gamma}^i_{00},~\tilde{\gamma}^0_{0i},~\tilde{\gamma}^i_{jk}$. Here, $|GM/r|$ denotes the typical value of Newtonian potential. Deducing from the definition of $l^\mu$, we find that $l^0$ is the same order with $1$ and $l^i$ is the same order with $GM/r$. Therefore, we obtain the approximation formula of $G^\mu$
\begin{eqnarray}\label{spray1}
G^0&=&2\tilde{\gamma}^0_{0i}y^0 y^i-l^0\tilde{b}_i(\tilde{\gamma}^i_{00}y^0 y^0+\tilde{\gamma}^i_{jk}y^j y^k),\\
\label{spray2}
G^i&=&\tilde{\gamma}^i_{00} y^0 y^0+\tilde{\gamma}^i_{jk}y^j y^k.
\end{eqnarray}
In Finsler geometry, one can get the Chern connection from the geodesic spray coefficient $G^\mu$\cite{Book by Bao}
\begin{equation}
N^\mu_\nu=\frac{1}{2}\frac{\partial G^\mu}{\partial y^\nu},~~\Gamma^\mu_{\nu\lambda}=\frac{\partial N^\mu_\nu}{\partial y^\lambda}+\frac{1}{2}g^{\mu\rho}y_\kappa\frac{\partial^2 N^\kappa_\rho}{\partial y^\nu \partial y^\lambda}.
\end{equation}
By making use of the approximation formula of $G^\mu$ (\ref{spray1}) and (\ref{spray2}), we obtain the approximation formula of the non vanish Chern connection
\begin{equation}
\Gamma^i_{00}=\tilde{\gamma}^i_{00},~\Gamma^0_{0i}=\tilde{\gamma}^0_{0i},~\Gamma^i_{jk}=\tilde{\gamma}^i_{jk}.
\end{equation}
Therefore, to first order ($GM/r$), the Chern connection is none other than the  Christoffel symbols of Riemannian metric $\tilde{a}$. It implies that curvature tensors approximately reduce to Riemannian, to first order ($GM/r$).  Following the deduction in \cite{Weinberg}, we obtain the result
\begin{equation}
\label{constrain}
\tilde{a}_{00}=-1+\delta a_{00},~~\tilde{a}_{ij}=\delta_{ij}(1+\delta a_{00}),
\end{equation}
where $\delta a_{00}$ is a perturbation term with order of $GM/r$.

\subsection{The Zermelo navigation problem}
Zermelo aimed to find minimum time trajectories in a Riemannian manifold ($M,h$), which under the influence of a wind represented by a vector field $W$\cite{Zermelo}. Shen \cite{Shen} proved that the minimum time trajectories are exactly the geodesic of a particular Finsler geometry-Randers space\cite{Randers}, if the wind is time independent. As a particular case of Finsler geometry, the geometrical properties of Randers space have been studied \cite{Bao}. Recently, Randers space has drawn physicists's attention. Gibbons {\it et al}. described connection between the Randers spacetime and the Zermelo viewpoints by casting the former in a Painlev\'{e}-Gullstrand form\cite{Gibbons}. It is interesting to study a particle moving in Randers-Finsler spacetime.

In the following, we study the Zermelo navigation problem which is regarded as an alternative of Randers-Finsler spacetime. For a given Riemannian metric $h$ and a wind $W$, one can set up a connection between Riemannnian data $(h,W)$ and Randers-Finsler structure $F$,
\begin{equation}\label{navigation data}
\tilde{a}_{ij}=\frac{\lambda h_{ij}+W_i W_j}{\lambda^2},~\tilde{b}_i=-\frac{W_i}{\lambda},~\lambda=1-h_{ij}W^i W^j,
\end{equation}
where $W^i=h^{ij}W_j$ and $\tilde{a}^{ij}=\lambda(h^{ij}-W^i W^j)$.

Schwarzschild metric as an exact solution of Einstein's vacuum field equation has been used to study four classical tests of general relativity. However, the problems we mentioned in the beginning of the paper can not be solved in the framework of Schwarzschild spacetime, if the gravitational source only involves baryonic matters. Here, we investigate the equation of motion in Finsler spacetime, which may be regarded as a candidate to solve the problems.

First, we start with the space part of Schwarzschild metric
\begin{equation}
h_{ij}dx^i dx^j=\left(1-\frac{2GM}{r}\right)^{-1}dr^2+r^2d\theta^2+r^2\sin\theta^2d\varphi^2,
\end{equation}
and consider the influence of a radial wind $W(r)\equiv W_r dr$. Then, the solution of Zermelo navigation problem gives the Randers metric, whose Finsler structure is given as
\begin{eqnarray}
Fd\tau&=&\sqrt{\lambda^{-1}\left(\left(1-\frac{2GM}{r}\right)^{-1}dr^2+r^2d\theta^2+r^2\sin\theta^2d\varphi^2\right)+\lambda^{-2}W_r^2dr^2}-\lambda^{-1}W_rdr\nonumber\\
\label{space part}
 &=&\sqrt{\lambda^{-2}\left(1-\frac{2GM}{r}\right)^{-1}dr^2+\lambda^{-1}(r^2d\theta^2+r^2\sin\theta^2d\varphi^2)}-\lambda^{-1}W_rdr,
\end{eqnarray}
where we have used the formula of $\lambda$ (\ref{navigation data}) to get the second equation of (\ref{space part}).

Next, we extend the structure (\ref{space part}) into relativistic form. By making use of the approximate solution (\ref{constrain}) of Finslerian vacuum field equation, we find that the Randers-Finsler structure is of this form
\begin{equation}\label{final form}
Fd\tau=\sqrt{-\lambda^2\left(1-\frac{2GM}{r}\right)dt^2+\lambda^{-2}\left(1-\frac{2GM}{r}\right)^{-1}dr^2+\lambda^{-1}(r^2d\theta^2+r^2\sin\theta^2d\varphi^2)}-\lambda^{-1}W_rdr,
\end{equation}
while $|GM/r|\ll1$.

The Schwarzschild metric is time independent and spatial isotropy, it means that there are Killing vectors which correspond to the conserve quantities $p_t=\left(1-\frac{2GM}{r}\right)\frac{dt}{d\tau}$ and $p_\varphi=r^2\sin^2\theta\frac{d\phi}{d\tau}$.  The Killing equations of Randers space are given as \cite{Finsler PF}
\begin{eqnarray}\label{k1}
K_V(\alpha)&=&\frac{1}{2\alpha}(V_{\mu|\nu}+V_{\nu|\mu})y^\mu y^\nu,\\
\label{k2}
K_V(\beta)&=&\left(V^\mu\frac{\partial \tilde{b}_\nu}{\partial x^\mu}+\tilde{b}_\mu\frac{\partial V^\mu}{\partial x^\nu}\right)y^\nu,
\end{eqnarray}
where $``|"$ denotes the covariant derivative with respect to the Riemannian metric $\alpha$. It is obvious that the vectors $V^0=C^0$ and $V^\varphi=C^\varphi$ are solutions of (\ref{k1}) in Rander-Finsler spacetime with structure (\ref{final form}), where $C^0$ and $C^\varphi$ are constant. Such solutions $V^0=C^0$ and $V^3=C^\varphi$ are also satisfy the equation (\ref{k2}). Thus, $V^0=C^0$ and $V^3=C^\varphi$ are Killing vectors in Rander-Finsler spacetime with structure (\ref{final form}). This symmetry is the same with Schwarzschild spacetime, it implies that conserve quantities exist in Rander-Finsler spacetime with structure (\ref{final form}), like $p_t$ and $p_\varphi$ in  Schwarzschild spacetime.

\subsection{The equation of motion}
One should notice that the Rander-Finsler structure (\ref{final form}) got in above subsection depends only on $r,dx^\mu$. Therefore, $\tilde{b}_\mu dx^\mu$ is a closed 1-form, the geodesic equation in such spacetime (\ref{final form}) is of the form
\begin{equation}\label{geodesic randers}
\frac{d^2x^\mu}{d\tau^2}+(\tilde{\gamma}^\mu_{\nu\lambda}+l^\mu\tilde{b}_{\nu|\lambda})\frac{dx^\nu}{d\tau}\frac{dx^\nu}{d\tau}=0.
\end{equation}
It is convenient to denote $B(r)\equiv\lambda^2\left(1-\frac{2GM}{r}\right)$ and $A(r)\equiv\lambda^{-2}\left(1-\frac{2GM}{r}\right)^{-1}$. By making use of the nonvanishing components of the connection $\tilde{\gamma}^\mu_{\nu\lambda}$, we find from (\ref{geodesic randers}) that
\begin{eqnarray}\label{eq t}
0&=&\frac{d^2t}{d\tau^2}+\frac{B'}{B}\frac{dt}{d\tau}\frac{dr}{d\tau}+\frac{dt}{d\tau}f\left(x,\frac{dx}{d\tau}\right),\\
\label{eq r}
0&=&\frac{d^2r}{d\tau^2}+\frac{A'}{2A}\left(\frac{dr}{d\tau}\right)^2+\frac{B'}{2A}\left(\frac{dt}{d\tau}\right)^2-\frac{1}{2A}\left(\frac{d\theta}{d\tau}\right)^2\frac{d}{dr}\left(\frac{r^2}{\lambda}\right)-\frac{\sin^2\theta}{2A}\left(\frac{d\varphi}{d\tau}\right)^2\frac{d}{dr}\left(\frac{r^2}{\lambda}\right)+\frac{dr}{d\tau}f\left(x,\frac{dx}{d\tau}\right),\\
\label{eq theta}
0&=&\frac{d^2\theta}{d\tau^2}+\frac{\lambda}{r^2}\frac{d}{dr}\left(\frac{r^2}{\lambda}\right)\frac{d\theta}{d\tau}\frac{dr}{d\tau}-\sin\theta\cos\theta\left(\frac{d\varphi}{d\tau}\right)^2+\frac{d\theta}{d\tau}f\left(x,\frac{dx}{d\tau}\right),\\
\label{eq varphi}
0&=&\frac{d^2\varphi}{d\tau^2}+\frac{\lambda}{r^2}\frac{d}{dr}\left(\frac{r^2}{\lambda}\right)\frac{d\varphi}{d\tau}\frac{dr}{d\tau}+2\cot\theta\frac{d\varphi}{d\tau}\frac{d\theta}{d\tau}+\frac{d\varphi}{d\tau}f\left(x,\frac{dx}{d\tau}\right),
\end{eqnarray}
where $f\left(x,\frac{dx}{d\tau}\right)\equiv\tilde{b}_{\nu|\lambda}\frac{dx^\nu}{d\tau}\frac{dx^\nu}{d\tau}/F$ and a prime denotes $d/dr$. Furthermore, we know that structure (\ref{final form}) only depends on $r,dx^\mu$. Thus, the field is isotropic.
It is convenient to consider the orbit of particle confined to the equatorial plane $\theta=\pi/2$. Then, the equation (\ref{eq theta}) is satisfied. Solving the equations (\ref{eq t}) and (\ref{eq varphi}), we get
\begin{eqnarray}\label{eq t1}
\frac{d}{d\tau}\left(\ln\frac{dt}{d\tau}+\ln B\right)&=&\frac{d\ln J_1}{d\tau},\\
\label{eq varphi1}
\frac{d}{d\tau}\left(\ln\frac{d\varphi}{d\tau}+\ln\frac{r^2}{\lambda}\right)&=&\frac{d\ln J_1}{d\tau},
\end{eqnarray}
where we involved a new quantity $J_1$. It is defined as$\frac{d\ln J_1}{d\tau}\equiv-f\left(x,\frac{dx}{d\tau}\right)$.
Deducing from the equations (\ref{eq t1}) and (\ref{eq varphi1}), we obtain two constants $(E,J)$ of motion. It satisfies
\begin{eqnarray}\label{eq motion1}
\frac{B}{J_1}\frac{dt}{d\tau}&=&E,\\
\label{eq motion2}
\frac{r^2}{\lambda J_1}\frac{d\varphi}{d\tau}&=&J.
\end{eqnarray}
By multiplying the equation (\ref{eq r}) with $2Adr/d\tau$ and making use of the equations (\ref{eq motion1},\ref{eq motion2}), we obtain
\begin{equation}\label{eq motion3}
B\left(\frac{dt}{d\tau}\right)^2-A\left(\frac{dr}{d\tau}\right)^2-\frac{r^2}{\lambda}\left(\frac{d\varphi}{d\tau}\right)^2=CJ_1^2,
\end{equation}
where $C$ is a constant.

In the following, the value of $C$ and the formula of $J_1$ will be derived. The derivative of the term $\tilde{b}_\mu\frac{dx^\mu}{d\tau}$ gives
\begin{eqnarray}
\frac{d}{d\tau}\left(\tilde{b}_\mu\frac{dx^\mu}{d\tau}\right)&=&\frac{dx^\nu}{d\tau}\frac{\partial}{\partial x^\nu}\left(\tilde{b}_\mu\frac{dx^\mu}{d\tau}\right)=\frac{dx^\nu}{d\tau}\left(\tilde{b}_\mu\frac{dx^\mu}{d\tau}\right)_{|\nu}\nonumber\\
&=&\tilde{b}_{\alpha|\beta}\frac{dx^\alpha}{d\tau}\frac{dx^\beta}{d\tau}+\tilde{b}_\mu\left(\frac{d^2x^\mu}{d\tau^2}+\tilde{\gamma}^\mu_{\nu\lambda}\right)\frac{dx^\nu}{d\tau}\frac{dx^\nu}{d\tau}\nonumber\\
\label{deduce J1}
&=&\left(1-\frac{\tilde{b}_\mu}{F}\frac{dx^\mu}{d\tau}\right)\tilde{b}_{\alpha|\beta}\frac{dx^\alpha}{d\tau}\frac{dx^\beta}{d\tau},
\end{eqnarray}
where $``|"$ denotes the covariant derivative with respect to the Riemannian metric $\alpha$.
Here, we have used the fact that the term $\tilde{b}_\mu\frac{dx^\mu}{d\tau}$ is a scaler in Riemannian spacetime with metric $\tilde{a}_{\mu\nu}$, to get the second equation of (\ref{deduce J1}). And we have used the geodesic equation (\ref{geodesic randers}) to get the last equation of (\ref{deduce J1}). Noticing that $F$ is constant along the geodesic, we find from equation (\ref{deduce J1}) that
\begin{equation}
\frac{d\ln\left(F-\tilde{b}_\mu\frac{dx^\mu}{d\tau}\right)}{d\tau}=-\tilde{b}_{\nu|\lambda}\frac{dx^\nu}{d\tau}\frac{dx^\nu}{d\tau}/F=-f\left(x,\frac{dx}{d\tau}\right).
\end{equation}
It implies that
\begin{equation}\label{formula J1}
J_1=F-\tilde{b}_\mu\frac{dx^\mu}{d\tau},
\end{equation}
with normalization of $\tau$. Combining the equations (\ref{eq motion3}) and (\ref{formula J1}), we find $C=1$ for massive particles. The equations (\ref{eq motion3}) and (\ref{formula J1}) can not determine the value of $C$ for photons. While $\tilde{b}$ vanishes, the equation of motion in Randers-Finsler spacetime must return to the one in Riemannian spacetime. This physical requirement implies $C=0$ for photons.

At last, we list what we got. In Rander-Finsler spacetime with structure (\ref{final form}), there are two constants of motion $E,J$, which again support our discussion about Killing vectors in subsection B. There is equation of motion (\ref{eq motion3}) corresponding to the constancy of $F$. All equations (\ref{eq motion1},\ref{eq motion2},\ref{eq motion3}) involve the term $J_1$, which could be regarded as deviation from the equation of motion in Schwarzschild spacetime. The constant $C$ in equation of motion equals 1 or 0, corresponds to massive particles and photons, respectively. Start with these results, we could find the trajectory of particles moving in Randers-Finsler spacetime.

\subsection{The Newtonian limit and gravitational deflection of light}
In the above subsection, we derived the equation of motion for particles in Randers-Finsler spacetime. In the following, we will study the orbit of particles. Recently, Grumiller constructed an effective model for gravity of a central object at large scales\cite{Grumiller}. It predicted a Rindler-type acceleration \cite{Wald}. We will show that if the Finslerian parameter $\lambda$ is of the form
\begin{equation}\label{Rindler lambda}
\lambda=1+\frac{GM}{r_s^2}r,
\end{equation}
it will deduce a Rindler-type acceleration,
where $r_s$ is a constant which denotes the physical scale of gravitational system.
First, we investigate the Newtonian limit. By making use of the equations (\ref{eq motion1},\ref{eq motion2},\ref{eq motion3}), we obtain the relation between the radial distant $r$ and time $t$
\begin{equation}\label{newton limit}
\frac{AE^2}{B^2}\left(\frac{dr}{dt}\right)^2+\frac{J^2\lambda}{r^2}-\frac{E^2}{B}=-1~.
\end{equation}
In Newtonian limit, particles move slowly in a weak field. Thus, the quantities $\frac{J^2}{r^2}, \left(\frac{dr}{dt}\right)^2, E^2-1, \frac{GM}{r}$ all are small. And to first order in these quantities, the equation (\ref{newton limit}) reduces to
\begin{equation}\label{newton limit1}
\frac{1}{2}\left(\frac{dr}{dt}\right)^2+\frac{J^2}{2r^2}-\left(1-\frac{r^2}{r_s^2}\right)\frac{GM}{r}=\varepsilon,
\end{equation}
where $\varepsilon\equiv\frac{1}{2}(E^2-1)$ is the energy in Newton's theory. The equation (\ref{newton limit1}) implies the effective Newtonian potential is modified
\begin{equation}\label{modified potential}
\phi_M=-\left(1-\frac{r^2}{r_s^2}\right)\frac{GM}{r}~.
\end{equation}
Therefore, the effective acceleration is Rindler-type acceleration
\begin{equation}\label{modified acc}
a_M=-\nabla\phi_M=-\left(\frac{GM}{r^2}+\frac{GM}{r_s^2}\right)~.
\end{equation}
Then, the velocity of galaxy rotational curve have the approximate relation
\begin{equation}\label{velocity}
v\approx\sqrt{\frac{GM}{r}+\frac{GM}{r_s^2}r}~.
\end{equation}
For a dwarf galaxies with mass scale $10^8$ solar masses, and $r_s=1$kiloparsec (a typical length scale of dwarf galaxies), we find that the term $\frac{GM}{r_s^2}$ in (\ref{velocity}) equals $10^{-62}$ in natural units ($c=\hbar=G=1$). For a spiral galaxies with mass scale $10^{11}$ solar masses, and $r_s=30$kiloparsec (a typical length scale of spiral galaxies), we find that the term $\frac{GM}{r_s^2}$ in (\ref{velocity}) also equals $10^{-62}$ in natural units ($c=\hbar=G=1$). Therefore, for a reasonable parameter $r_s$, our result is compatible with Grumiller's results of galaxies rotational curve \cite{Grumiller}. It means that our result (\ref{velocity}) could account for galaxies rotational curve.

Next, we investigate the gravitational deflection of light. We limit us to the case of weak gravitational field. It implies $\frac{GM}{r}\ll1$. By making use of the equations (\ref{eq motion1},\ref{eq motion2},\ref{eq motion3}), and noticing that $C=0$ for photons, we obtain the relation between the radial distant $r$ and angle $\varphi$
\begin{eqnarray}
\left(\frac{1}{r^2}\frac{dr}{d\varphi}\right)^2&=&\left(\frac{E}{J}\right)^2\frac{1}{AB\lambda^2}-\frac{1}{Ar^2\lambda}\nonumber\\
\label{bend light}
&=&\left(\frac{E}{J\lambda}\right)^2-\frac{\lambda}{r^2}\left(1-\frac{2GM}{r}\right).
\end{eqnarray}
In terms of variable $u=\frac{GM}{r}$, the equation (\ref{bend light}) changes as
\begin{equation}\label{eq u}
\left(\frac{du}{d\varphi}\right)^2=\left(\frac{EGM}{J\lambda}\right)^2-\lambda u^2(1-2u).
\end{equation}
 At the closest approach to the gravitational source $M$, $\varphi=\varphi_m,~u=u_m,~\lambda=\lambda_m$ and $du/d\varphi$ vanishes, thus
\begin{equation}\label{singularity}
\left(\frac{EGM}{J}\right)^2=\lambda_m^3u_m^2(1-2u_m).
\end{equation}
Substituting the equation (\ref{singularity}) into (\ref{eq u}), we obtain
\begin{equation}\label{singularity1}
\frac{d\varphi}{du}=\frac{\lambda}{\sqrt{\lambda_m^3u_m^2(1-2u_m)-\lambda^3u^2(1-2u)}}.
\end{equation}

Changing variable to $x=u/u_m$, to first order in $u_m$, we obtain from (\ref{Rindler lambda},\ref{singularity1}) that
\begin{eqnarray}
\varphi_m-\varphi_\infty&=&\int^1_0\frac{dx}{\sqrt{1-x^2-2u_m(1-x^3)+3u_s^2/u_m-2u_s^2/u_mx-xu_s^2/u_m}}\nonumber\\
                        &=&\int^1_0\frac{dx}{\sqrt{1-x^2}}\left(1+u_m\frac{1-x^3}{1-x^2}-\frac{3u_s^2/u_m-2u_s^2/u_mx-xu_s^2/u_m}{2(1-x^2)}\right)\nonumber\\
                        \label{singularity2}
                        &=&\frac{\pi}{2}+2u_m+\frac{u_s^2}{u_m}\left(\frac{3\sqrt{1-x^2}}{2(1+x)}-\log\left(\frac{1+\sqrt{1-x^2}}{x}\right)\right)\bigg|^1_0
\end{eqnarray}
The term $\frac{\pi}{2}+2u_m$ is what we expect in general relativity, and the third term in (\ref{singularity2}) tends to infinity at $x=0$. This infinity is due to that the Rindler potential (\ref{modified potential}) is infinity at infinity distance $(u=0)$. It is reasonable to consider that the Rindler-type potential vanishes outsider a given cut off scale $r_{\rm out}$.
Therefore, we deduce from (\ref{singularity2}) that the deflection angle for the Rindler-type acceleration in Finsler spacetime is of the form
\begin{equation}
\label{deflection Rindler}
\alpha(r_m)=\frac{GM}{r_m}\left(4+2\frac{r_m^2}{r_s^2}\left[\log\left(\frac{r_{\rm out}+\sqrt{r_{\rm out}^2-r_m^2}}{r_m}\right)-\frac{3\sqrt{r_{\rm out}^2-r_m^2}}{2(r_{\rm out}+r_m)}\right]\right).
\end{equation}
While $r_m\ll r_s$, the deflection angle (\ref{deflection Rindler}) reduces to the familiar one in general relativity.

The strong and weak gravitational lensing survey of Bullet cluster 1E0657-558\cite{Bullet} obtained a convergence $\kappa$-map. The convergence $\kappa$ is defined as\cite{Peacock}
\begin{equation}
\label{defing kappa}
\kappa=\frac{D_{LS}}{2D_S}\nabla_\theta\alpha(\theta_I)=\frac{D_{LS}D_L}{2D_S}\nabla_{\vec{\xi}}~\alpha(\vec{\xi}),
\end{equation}
where $\alpha(\vec{\xi})$ is deflection angle, $D_{LS}$ is the angular distance
from the lens plane to a source galaxy, $D_L$ is the angular distance to the lens plane, $D_S$ is the angular distance to a source
galaxy, $\theta_I$ specify the observed position of the source galaxy, $\vec{\xi}$ is the two dimensional vector in lens plane and $\nabla_{\vec{\xi}}$ is two dimensional gradient operator. The surface density $\Sigma(\vec{\xi})$ is derived as
\begin{equation}
\Sigma(\vec{\xi})=\int\rho(\vec{\xi},z)dz,
\end{equation}
where $\rho(\vec{\xi},z)$ is the density of gravitational source $M$ and $z$ is the direction perpendicular to the lens plane. In weak field , the deflection angle can be obtained as the superposition of the deflections
\begin{equation}\label{deflection kappa}
\alpha(\vec{\xi})=4G\int\bar{\Sigma}(\vec{\xi}')\frac{\vec{\xi}-\vec{\xi}'}{|\vec{\xi}-\vec{\xi}'|^2},
\end{equation}
where
\begin{equation}\label{convergence}
\bar{\Sigma}(\vec{\xi})=\Sigma(\vec{\xi})\left(1+\frac{|\vec{\xi}|^2}{2r_s^2}\left[\log\left(\frac{r_{\rm out}+\sqrt{r_{\rm out}^2-|\vec{\xi}|^2}}{|\vec{\xi}|}\right)-\frac{3\sqrt{r_{\rm out}^2-|\vec{\xi}|^2}}{2(r_{\rm out}+|\vec{\xi}|)}\right]\right).
\end{equation}
Substituting (\ref{deflection kappa}) into (\ref{defing kappa}), we obtain
\begin{equation}\label{convergence1}
\kappa=\frac{4\pi GD_{LS}D_L}{D_S}\bar{\Sigma}(\vec{\xi})\equiv\frac{\bar{\Sigma}(\vec{\xi})}{\Sigma_c}~.
\end{equation}
It is obvious from (\ref{convergence},\ref{convergence1}) that the convergence $\kappa$ does not reach its maximum at the center of $\Sigma(\xi)$. It also implies that the position where $\kappa$ reaches its maximum value is separated from the center of $\Sigma(\xi)$. Thus, the convergence $\kappa$ deduced in Finsler gravity satisfies the features of Bullet Cluster. The Rindler-type potential in Finsler spacetime could account for the observations of Bullet Cluster.

\section{Conclusions}
In this paper, we presented the vacuum field equation in Finsler spacetime. By making use of the post-Newtonian approximation and the viewpoints of Zermelo navigation problem, we investigated the dynamics in Randers-Finsler spacetime. The Newtonian limit and gravitational deflection of light was obtained explicitly. Within the framework of Finsler spacetime, the deflection angle and the convergence $\kappa$ in Rindler-type potential were given.

The surface density $\Sigma$-map and the convergence $\kappa$-map of Bullet Cluster 1E0657-558\cite{Bullet} show that the center of baryonic matters separate from the center of gravitational force, and the distribution of gravitational force do not possess spherical symmetry. The formula (\ref{modified acc}) manifests that the gravity in Finsler spacetime modified the Newtonian inverse-square law at large scale. It is obvious from (\ref{convergence},\ref{convergence1}) that the convergence $\kappa$ does not reach its maximum at the center of $\Sigma(\xi)$. It also implies that the position where $\kappa$ reaches its maximum value is separated from the center of $\Sigma(\xi)$. Thus, our model satisfies the first particular feature of Bullet Cluster. The particular feature of spherical symmetry broken implies that all modified gravity models with central potential need improvements for accounting the observations of Bullet Cluster. However, for simplicity, the central potential could regarded as the zero order term of the final modified gravity model for Bullet Cluster. The spherical symmetry broken may be deduced by the next leading order term of the final modified gravity model.

The convergence $\kappa$ in Rindler-type potential could account for observations of Bullet Cluster. The numerical analysis is in progress. In future work, we will consider the non-central potential in Finsler spacetime, and investigate the effect of spherical symmetry broken.

\vspace{1cm}
\begin{acknowledgments}
We would like to thank Prof. C. J. Zhu, M. H. Li and S. Wang for useful discussions. The
work was supported by the NSF of China under Grant No. 10875129 and 11075166.
\end{acknowledgments}

\end{document}